# Optimization with More than One Budget*

Fabrizio Grandoni†      Rico Zenklusen‡

November 13, 2018

## Abstract

A natural way to deal with multiple, partially conflicting objectives is turning all the objectives but one into budget constraints. Some classical polynomial-time optimization problems, such as spanning tree and forest, shortest path, (perfect) matching, independent set (basis) in a matroid or in the intersection of two matroids, become NP-hard even with one budget constraint. Still, for most of these problems deterministic and randomized polynomial-time approximation schemes are known. In the case of two or more budgets, typically only multi-criteria approximation schemes are available, which return slightly infeasible solutions. Not much is known however for the case of strict budget constraints: filling this gap is the main goal of this paper.

We show that shortest path, perfect matching, and spanning tree (and hence matroid basis and matroid intersection basis) are inapproximable already with two budget constraints. For the remaining problems, whose set of solutions forms an *independence system*, we present deterministic and randomized polynomial-time approximation schemes for a constant number $k$ of budget constraints. Our results are based on a variety of techniques:

1. We present a simple and powerful mechanism to transform multi-criteria approximation schemes into *pure* approximation schemes. This gives, for example, deterministic approximation schemes for $k$-budgeted forest and bipartite matching, and randomized approximation schemes for $k$-budgeted matching, independent set in matroids, and independent set in the intersection of two representable matroids.

2. We show that points in low dimensional faces of any matroid polytope are almost integral, an interesting result on its own. This gives a deterministic approximation scheme for $k$-budgeted matroid independent set.

3. We present a deterministic approximation scheme for 2-budgeted matching. The backbone of this result is a purely topological property of curves in $\mathbb{R}^2$.

## 1 Introduction

In many applications, one has to compromise between several, partially conflicting goals. *Multi-Objective Optimization* is a broad area of study in Operations Research, Economics and Computer Science [11, 14, 29]. A variety of approaches have been employed to formulate such problems including Goal Programming [4], Pareto-Optimality [10], and Multi-Budgeted Optimization [29]. We adopt the latter approach and cast one of the goals as the objective function, and the others as *budget constraints*. More precisely, we are given a (finite) set $\mathcal{F}$ of solutions for the problem, where each solution is a subset $S$ of elements from a given universe $E$ (e.g., the edges of a graph). We

---

are also given a weight function $w : \mathcal{F} \to \mathbb{Q}_+$ and a set of $k = O(1)^1$ length functions $\ell_i : \mathcal{F} \to \mathbb{Q}_+$, $1 \leq i \leq k$, that assign a weight $w(S) := \sum_{e \in S} w(e)$ and an $i$th-length $\ell_i(S) := \sum_{e \in S} \ell_i(e)$, $1 \leq i \leq k$, to every candidate solution $S$. For each length function $\ell_i$, there is a budget $L_i \in \mathbb{Q}_+$. The *multi-budgeted optimization problem* can then be formulated as follows:

$$\text{maximize/minimize } w(S) \text{ subject to } S \in \mathcal{F}, \ \ell_i(S) \leq L_i, \ 1 \leq i \leq k. \qquad (1)$$

We next use $OPT$ to denote an optimum solution.

Following the literature on the topic, we focused on the set of problems below:

- $k$-BUDGETED (PERFECT) MATCHING: $\mathcal{F}$ is given by the (perfect) matchings of an undirected graph $G = (V, E)$.
- $k$-BUDGETED SPANNING TREE (FOREST): $\mathcal{F}$ is given by the spanning trees (forests) of $G$.
- $k$-BUDGETED SHORTEST PATH: $\mathcal{F}$ is given by the paths connecting two given nodes $s$ and $t$ in $G$.
- $k$-BUDGETED MATROID INDEPENDENT SET (BASIS): $\mathcal{F}$ is given by the independent sets (bases) of a matroid $M = (E, \mathcal{I})^2$.
- $k$-BUDGETED MATROID INTERSECTION INDEPENDENT SET (BASIS): $\mathcal{F}$ is given by the independent sets (bases) in the intersection of two matroids $M_1 = (E, \mathcal{I}_1)$ and $M_2 = (E, \mathcal{I}_2)$.

All the problems above are polynomial-time solvable (see, e.g., [31]) in their unbudgeted version ($k = 0$), but become NP-hard [1, 6, 12] even for a single budget constraint ($k = 1$). For the case of one budget ($k = 1$), polynomial-time approximation schemes (PTASs) are known for SPANNING TREE [28] (see also [17]), SHORTEST PATH [32] (see also [16, 21]), and MATCHING [6] (see also [5]). The approach in [28] easily generalizes to the case of MATROID BASIS. A PTAS is also known for MATROID INTERSECTION INDEPENDENT SET [6]. The results in [6] do not generalize to the case of PERFECT MATCHING and MATROID INTERSECTION BASIS. No approximation algorithm is known for the problems above in the case $k \geq 2$ (excluding multi-criteria algorithms which provide slightly infeasible solutions): investigating the existence of such algorithms is the main goal of this paper.

## 1.1 Our Results

We start by observing that several of the mentioned problems are inapproximable already for two budget constraints. More precisely, the corresponding feasibility problem is NP-complete. The simple proof of the following theorem is given in the appendix.

**Theorem 1.** *For $k \geq 2$, unless $P = NP$ there is no polynomial-time approximation algorithm for $k$-BUDGETED SPANNING TREE, $k$-BUDGETED SHORTEST PATH, $k$-BUDGETED PERFECT MATCHING, $k$-BUDGETED MATROID BASIS, and $k$-BUDGETED MATROID INTERSECTION BASIS.*

The remaining problems have a common aspect: the set of solutions $\mathcal{F}$ forms an *independence system*. In other terms, for $S \in \mathcal{F}$ and $S' \subseteq S$, we have $S' \in \mathcal{F}$. For these problems, we present deterministic and randomized approximation schemes, based on a variety of techniques.

---

[1]The assumption that $k$ is a constant is crucial in this paper.

[2]We recall that $E$ is a finite ground set and $\mathcal{I} \subseteq 2^E$ is a nonempty family of subsets of $E$ (*independent sets*) which have to satisfy the following two conditions: (i) $I \in \mathcal{I}, J \subseteq I \Rightarrow J \in \mathcal{I}$ and (ii) $I, J \in \mathcal{I}, |I| > |J| \Rightarrow \exists z \in I \setminus J : J \cup \{z\} \in \mathcal{I}$. A *basis* is a maximal independent set. For all matroids used in this paper we make the usual assumptions that independence of a set can be checked in polynomial time. For additional information on matroids, see e.g. [31].

Our first result (see Section 2) is a simple but powerful mechanism to transform a multi-criteria PTAS, i.e. a PTAS that might violate the budgets by a small multiplicative factor, into a *pure* PTAS, where no budget is violated. Similarly, a multi-criteria polynomial randomized-time approximation scheme (PRAS) can be transformed into a pure PRAS.

**Theorem 2. (Feasibilization)** *Let $\mathcal{P}_{ind}$ be a $k$-budgeted problem where the set of solutions $\mathcal{F}$ is an independence system. Suppose that we are given an algorithm $\mathcal{A}$ which, for any constant $\delta > 0$, computes in polynomial time an $(1-\delta)$ (resp., expected $(1-\delta)$) approximate solution to $\mathcal{P}_{ind}$ violating each budget by a factor at most $(1+\delta)$. Then there is a PTAS (resp., PRAS) for $\mathcal{P}_{ind}$.*

The idea behind the proof is showing that a *good* solution exists even if we scale down the budgets by a small factor. This is done by applying a greedy discarding strategy similar to the greedy algorithm for KNAPSACK. Applying a multi-criteria PTAS (given as a black box!) to the scaled problem gives a feasible solution for the original one, of weight *close* to the optimal weight.

To the best of our knowledge, this simple result was never observed before. Indeed, it implies improved approximation algorithms for a number of problems. A general construction by Papadimitriou and Yannakakis [25] provides multi-criteria PTASs (resp., PRASs) for problems whose exact version admits a pseudo-polynomial-time (PPT) deterministic (resp., Monte-Carlo) algorithm. We recall that the *exact version* of a given optimization problem asks for a feasible solution of exactly a given *target* weight. Combining their approach with our mechanism one obtains approximation schemes for several problems. For example, using the PPT-algorithm for EXACT FOREST in [3], one obtains a PTAS for $k$-BUDGETED FOREST. Similarly, the Monte-Carlo PPT-algorithm for EXACT MATCHING in [24] gives a PRAS for $k$-BUDGETED MATCHING. The Monte-Carlo PPT-algorithms for EXACT MATROID INTERSECTION INDEPENDENT SET in [8], which works in the special case of representable matroids[3], implies a PRAS for the corresponding budgeted problem.

Of course, one can also exploit multi-criteria approximation schemes obtained with different techniques. For example, exploiting the multi-criteria PTAS in [13] for $k$-BUDGETED MATCHING in bipartite graphs, which is based on iterative rounding, one obtains a PTAS for the same problem. Very recently [9], a multi-criteria PRAS for $k$-BUDGETED MATROID INDEPENDENT SET, based on dependent randomized rounding, has been presented. This implies a PRAS for $k$-BUDGETED MATROID INDEPENDENT SET.

**Corollary 3.** *There are PTASs for $k$-BUDGETED FOREST and $k$-BUDGETED MATCHING in bipartite graphs. There are PRASs for $k$-BUDGETED MATCHING, $k$-BUDGETED MATROID INDEPENDENT SET, and $k$-BUDGETED MATROID INTERSECTION in representable matroids.*

Based on a different, more direct approach, we are able to turn the PRAS for $k$-BUDGETED MATROID INDEPENDENT SET into a PTAS. The main insight is the following structural property of faces of the matroid polytope which might be of independent interest (proof in Section 3).

**Theorem 4.** *Let $M = (E, \mathcal{I})$ be a matroid and let $F$ be a face of dimension $d$ of the matroid polytope[4] $P_{\mathcal{I}}$. Then any $x \in F$ has at most $2d$ non-integral components. Furthermore, the sum of all fractional components of $x$ is at most $d$.*

A PTAS can then be easily derived as follows. We first guess the $k/\varepsilon$ elements $E_H$ of largest weight in the optimum solution in a preliminary phase, and reduce the problem consequently. This guessing step guarantees that the maximum weight $w_{max}$ of an element in the reduced problem satisfies $kw_{max} \leq \varepsilon w(E_H)$. For the reduced problem, we compute an optimal fractional vertex solution $x^*$ to the LP which seeks to find a maximum weight point in the matroid polytope intersected with the $k$ budget constraints. Since $x^*$ is chosen to be a vertex solution, and only $k$ linear constraints are added to the matroid polytope, $x^*$ lies on a face of the matroid polytope of dimension at most $k$. We then round down the fractional components of $x^*$ to obtain an incidence vector $\overline{x}$ which corresponds to some independent set $E_L$. By Theorem 4, $|x^* - \overline{x}| \leq k$, and hence, $w(E_L) \geq w(x^*) - kw_{max}$. Then, it is not hard to see that $E_H \cup E_L$ is a $(1-\varepsilon)$-approximate feasible solution for the starting problem.

---
[3]A matroid $M = (S, \mathcal{I})$ is representable if its ground set $S$ can be mapped in a bijective way to the columns of a matrix over some field, and $I \subseteq S$ is independent in $M$ iff the corresponding columns are linearly independent.

[4]For some given matroid $M = (E, \mathcal{I})$, the corresponding matroid polytope $P_{\mathcal{I}}$ is the convex hull of the incidence vectors of all independent sets.

**Corollary 5.** *There is a PTAS for $k$-BUDGETED MATROID INDEPENDENT SET.*

Eventually, we present a PTAS (rather than a PRAS as in Corollary 3) for 2-BUDGETED MATCHING (see Section 4).

**Theorem 6.** *There is a PTAS for 2-BUDGETED MATCHING.*

Our PTAS, works as follows. Let us confuse a matching $M$ with the associated incidence vector $x_M$. We initially compute an optimal fractional matching $x^*$, and express it as the convex combination $x^* = \alpha_1 x_1 + \alpha_2 x_2 + \alpha_3 x_3$ of three matchings $x_1$, $x_2$, and $x_3$. Then we exploit a *patching procedure* which, given two matchings $x'$ and $x''$ with high Lagrangian weight and a parameter $\mu \in [0, 1]$, computes a matching $z$ which is not longer than $x_\mu := \mu x' + (1-\mu)x''$ with respect to both lengths, and has a comparable weight. This procedure is applied twice: first on the matchings $x_1$ and $x_2$ with parameter $\mu = \alpha_1/(\alpha_1 + \alpha_2)$, hence getting a matching $z'$. Second, on the two matchings $z'$ and $x_3$ with parameter $\mu = (\alpha_1 + \alpha_2)/(\alpha_1 + \alpha_2 + \alpha_3)$. The resulting matching $z''$ is feasible and almost optimal (modulo a preliminary guessing step).

Our patching procedure relies on a topological property of curves in $\mathbb{R}^2$, that we prove via Jordan's curve theorem [22]. An extension of the property above to curves in $\mathbb{R}^k$ would imply a PTAS for $k$-BUDGETED MATCHING: this is left as an interesting open problem.

### 1.2 Related Work

There are a few general tools for designing approximation algorithms for budgeted problems. One basic approach is combining *dynamic programming* (which solves the problem for polynomial weights and lengths) with *rounding and scaling* techniques (to reduce the problem to the case of polynomial quantities). This leads for example to the FPTAS for 1-BUDGETED SHORTEST PATH [16, 21, 32]. Another fundamental technique is the *Lagrangian relaxation method*. The basic idea is relaxing the budget constraints, and lifting them into the objective function, where they are weighted by Lagrangian multipliers. Solving the relaxed problem, one obtains two or more solutions with optimal Lagrangian weight, which can - if needed - be patched together to get a good solution for the original problem. Demonstrating this method, Goemans and Ravi [28] gave a PTAS for 1-BUDGETED SPANNING TREE, which also extends to 1-BUDGETED MATROID BASIS. Using the same approach, with an involved patching step, Berger, Bonifaci, Grandoni, and Schäfer [6] obtained a PTAS for 1-BUDGETED MATCHING and 1-BUDGETED MATROID INTERSECTION INDEPENDENT SET. Their approach does not seem to generalize to the case of multiple budget constraints.

The techniques above apply to the case of one budget. Not much is known for problems with two or more budgets. However, often *multi-criteria* approximation schemes are known, which provide a $(1-\varepsilon)$-approximate solution violating the budgets by a factor $(1+\varepsilon)$. First of all, there is a very general technique by Papadimitriou and Yannakakis [25], based on the construction of $\varepsilon$-approximate Pareto curves. Given an optimization problem with multiple objectives, the *Pareto* *curve* consists of the set of solutions $S$ such that there is no solution $S'$ which is strictly better than $S$ (in a vectorial sense). Papadimitriou and Yannakakis show that, for any constant $\varepsilon > 0$, there always exists a polynomial-size $\varepsilon$-approximate Pareto curve $\mathcal{A}$, i.e., a set of solutions such that every solution in the Pareto curve is within a factor of $(1 + \varepsilon)$ from some solution in $\mathcal{A}$ on each objective. Furthermore, this approximate curve can be constructed in polynomial time in the size of the input and $1/\varepsilon$ whenever there exists a PPT algorithm for the associated exact problem. This implies multi-criteria FPTASs for $k$-BUDGETED SPANNING TREE and $k$-BUDGETED SHORTEST PATH. Furthermore, it implies a multi-criteria FPRAS for $k$-BUDGETED (PERFECT) MATCHING. The latter result exploits the Monte-Carlo PPT algorithm for EXACT MATCHING in [24]. Our PRAS improves on these results, approximation-wise (the running time is larger in our case).

Recently, Grandoni, Ravi and Singh [13] showed that the *iterative rounding* technique is an alternative way to achieve similar (or better) results. The idea behind iterative rounding [18] (see also, e.g., [2, 20]) is to consider a linear relaxation of the problem, compute an optimal fractional solution, and round one of its variables. The process is then iterated on the residual problem until a feasible integral solution is obtained. This approach can be enhanced with a relaxation step, where a constraint which cannot be violated too much is relaxed (i.e., deleted). Using this method, Grandoni et al. obtain a multi-criteria PTAS for $k$-BUDGETED SPANNING TREE, which computes a solution of optimal cost violating each budget by a factor $(1 + \varepsilon)$. This improves, approximation-wise, on the result in [25] for the same problem (where the solution returned is suboptimal). The authors also show how to obtain a deterministic (rather than randomized [25]) multi-criteria PTAS for $k$-BUDGETED MATCHING in bipartite graphs.

All the mentioned problems are easy in the unbudgeted version. Given an NP-hard unbudgeted problem which admits a $\rho$ approximation, the *parametric search* technique in [23] provides a multi-criteria $k\rho$ approximation algorithm violating each budget by a factor $k\rho$ for the corresponding problem with $k$ budgets. Other techniques lead to logarithmic approximation factors (see, e.g., [7, 26, 27]).

## 2 A Feasibilization Mechanism

In this section we illustrate our feasibilization mechanism, proving Theorem 2.

*Proof of Theorem 2.* Let $\varepsilon \in (0,1]$ be a given constant. Consider the following algorithm. Initially we guess the $h = k/\varepsilon$ elements[5] $E_H$ of $OPT$ of largest weight, and reduce the problem consequently[6], hence getting a problem $\mathcal{P}'$. Then we scale down all the budgets by a factor $(1 - \delta)$, and solve the resulting problem $\mathcal{P}''$ by means of $\mathcal{A}$, where $\delta = \varepsilon/(k+1)$. Let $E_L$ be the solution returned by $\mathcal{A}$. We eventually output $E_H \cup E_L$.

Let $OPT'$ and $OPT''$ be the optimum solution to problems $\mathcal{P}'$ and $\mathcal{P}''$, respectively. We also denote by $L'_i$ and $L''_i$ the $i$th budget in the two problems, respectively. Eventually, let $w_{max}$ be the largest weight in $\mathcal{P}'$ and $\mathcal{P}''$. We observe that trivially

$$(a)\ w(OPT) = w(E_H) + w(OPT') \qquad \text{and} \qquad (b)\ w_{max} \leq w(E_H)/h.$$

Let us show that
$$w(OPT'') \geq w(OPT')(1 - k\delta) - kw_{max}. \tag{2}$$

---
[5]To avoid inessential technicalities, we assume $1/\varepsilon \in \mathbb{N}$.
[6]As usual, by reducing we mean decreasing each budget $L_i$ by $\ell_i(E_H)$ and removing all the elements of weight strictly larger than $\min_{e \in E_H} w(e)$. By guessing we mean trying all the $O(m^h)$ subsets of $h$ elements.

Consider the following process: for each length function $i$, we remove from $OPT'$ the element $e$ with smallest ratio $w(e)/\ell_i(e)$ until $\ell_i(OPT') \leq (1-\delta)L'_i$. Let $E_i$ be the set of elements removed. It is not hard to see that $w(E_i) \leq \delta w(OPT') + w_{max}$. It follows that $OPT' - \cup_i E_i$ is a feasible solution for $\mathcal{P}''$ of weight at least $w(OPT')(1-\delta k) - kw_{max}$, proving (2).

We observe that $E_L$ is feasible for $\mathcal{P}'$ since, for each $i$,
$$\ell_i(E_L) \leq (1+\delta)L''_i = (1+\delta)(1-\delta)L'_i \leq L'_i.$$

As a consequence, the returned solution $E_H \cup E_L$ is feasible. Moreover, when $\mathcal{A}$ is deterministic, we have

$$w(E_H) + w(E_L) \geq w(E_H) + (1-\delta)w(OPT'') \stackrel{(2)}{\geq} w(E_H) + (1-\delta)(w(OPT')(1-\delta k) - kw_{max})$$
$$\stackrel{(b)}{\geq} (1-k/h)w(E_H) + (1-\delta(k+1))w(OPT') \geq (1-\varepsilon)(w(E_H) + w(OPT'))$$
$$\stackrel{(a)}{=} (1-\varepsilon)w(OPT).$$

The same bound holds in expectation when $\mathcal{A}$ is randomized. $\square$

## 3 A PTAS for $k$-Budgeted Matroid Independent Set

It is convenient to consider weights w and lengths $\ell_i$ as vectors in $\mathbb{Q}^E$. We denote by $\ell$ the matrix whose $i$th column is $\ell_i$, and let $L = (L_1, \ldots, L_k)^T$. To every matroid $M = (E, \mathcal{I})$, a *rank function* $r : 2^E \to \mathbb{N}$ is associated; it is defined by $r(S) = \max\{|J| \mid J \subseteq S, J \in \mathcal{I}\}$. The *matroid polytope* $P_\mathcal{I}$ is the convex hull of the characteristic vectors $\chi_I$ of the independent sets $I \in \mathcal{I}$ and is described by the following set of inequalities (see [31] for more details):

$$P_\mathcal{I} = \text{conv}\{\chi_I : I \in \mathcal{I}\} = \{x \geq 0 : x(S) \leq r(S) \; \forall S \subseteq E\}.$$

*Proof of Theorem 4.* Let $m = |E|$. We assume that the matroid polytope has full dimension, i.e., $\dim(P_\mathcal{I}) = m$, which is equivalent to saying that every element $e \in E$ is independent. This can be assumed without loss of generality since if $\{e\} \notin \mathcal{I}$ for some $e \in E$, then we can reduce the matroid by deleting element $e$. Since $\dim(P_\mathcal{I}) = m$ and $\dim(F) = d$, $F$ can be described by the inequality system of $P_\mathcal{I}$, where $m - d$ linearly independent inequalities used in the description of $P_\mathcal{I}$ are turned into equalities. More precisely, there are $N \subseteq E$ and $A_1, \ldots, A_k \subseteq E$ such that

$$F = \{x \in P_\mathcal{I} \mid x(e) = 0 \; \forall e \in N, x(A_i) = r(A_i) \; \forall i \in \{1, \ldots, k\}\},$$

and $|N| + k = m - d$. By standard uncrossing arguments, we can assume that the sets $A_i$ form a chain, i.e., $A_1 \subsetneq A_2 \subsetneq \cdots \subsetneq A_k$ (see for example [15, 18] for further information on combinatorial uncrossing). We prove the claim by induction on the number of elements of the matroid. The theorem clearly holds for matroids with a ground set of cardinality one. First assume $N \neq \emptyset$ and let $e \in N$. Let $M'$ be the matroid obtained from $M$ by deleting $e$, and let $F'$ be the projection of $F$ onto the coordinates corresponding to $N \setminus \{e\}$. Since $F'$ is a face of $M'$, the claim follows by induction. Henceforth, we assume $N = \emptyset$ which implies $k = m - d$. Let $A_0 = \emptyset$ and $B_i = A_i \setminus A_{i-1}$ for $i \in \{1, \ldots, k\}$. In the following we show that we can assume

$$0 < r(A_i) - r(A_{i-1}) < |B_i| \quad \forall \, i \in \{1, \ldots, k\}. \tag{3}$$

Notice that $0 \leq r(A_i) - r(A_{i-1}) \leq |B_i|$ clearly holds by standard properties of rank functions (see [31] for more details). Assume that there is $i \in \{1, \ldots, k\}$ with $r(A_i) = r(A_{i-1})$. Since all points $x \in F$ satisfy $x(A_i) = r(A_i)$ and $x(A_{i-1}) = r(A_{i-1})$, we have $x(B_i) = 0$. Hence for any $e \in B_i$, we have $x(e) = 0$ for $x \in F$. Again, we can delete $e$ from the matroid, hence obtaining a smaller matroid for which the claim holds by the inductive hypothesis. Therefore, we can assume $r(A_i) > r(A_{i-1})$ which implies the left inequality in (3).

For the right inequality assume that there is $i \in \{1, \ldots, k\}$ with $r(A_i) - r(A_{i-1}) = |B_i|$. Hence, every $x \in F$ satisfies $x(B_i) = |B_i|$, implying $x(e) = 1$ for all $e \in B_i$. Let $e \in B_i$, and let $F'$ be the projection of the face $F$ onto the components $N \setminus \{e\}$. Since $F'$ is a face of the matroid $M'$ obtained from $M$ by contracting $e$, the result follows again by the inductive hypothesis.

Henceforth, we assume that (3) holds. This implies in particular that $|B_i| > 1$ for $i \in \{1, \ldots, k\}$. Since $\sum_{i=1}^{k} |B_i| \leq m$, we have $k \leq m/2$, which together with $k = m - d$ implies $d \geq m/2$. The claim of the theorem that $x \in F$ has at most $2d$ non-integral components is thus trivial in this case. To prove the second part of the theorem we show that if (3) holds then $x(E) \leq d$ for $x \in F$. For $x \in F$ we have

$$x(E) = x(E \setminus A_k) + \sum_{i=1}^{k} x(B_i) \leq |E| - |A_k| + \sum_{i=1}^{k}(r(A_i) - r(A_{i-1}))$$

$$\leq |E| - |A_k| + \sum_{i=1}^{k}(|A_i| - |A_{i-1}| - 1) = m - k = d,$$

where the first inequality follows from $x(E \setminus A_k) \leq |E \setminus A_k|$ and $x(B_i) = r(A_i) - r(A_{i-1})$, and the second inequality follows from (3). $\square$

## 4 A PTAS for 2-BUDGETED MATCHING

In this section we present our PTAS for 2-BUDGETED MATCHING. We denote by $\mathcal{M}$ the set of incidence vectors of matchings. With a slight abuse of terminology we call the elements in $\mathcal{M}$ matchings. Let $P_\mathcal{M}$ be the matching polyhedron. Analogously to Section 3, let $\ell = (\ell_1, \ell_2)$ and $L = (L_1, L_2)^T$. A feasible solution in this framework is a matching $x \in \mathcal{M}$ such that $\ell^T x \leq L$. For two elements $z', z'' \in [0,1]^E$, we define their *symmetric difference* $z' \Delta z'' \in [0,1]^E$ by $(z' \Delta z'')(e) = |z'(e) - z''(e)|$ for all $e \in E$. In particular, if $z'$ and $z''$ are incidence vectors, then their symmetric difference as defined above corresponds indeed to the symmetric difference in the usual sense. Recall that, when $z'$ and $z''$ are matchings, $z' \Delta z''$ consists of a set of node-disjoint paths and cycles.

We start by presenting a property of curves in $\mathbb{R}^2$ (Section 4.1). This property is used to derive the mentioned *patching procedure* (Section 4.2). Eventually, we describe and analyze our PTAS (Section 4.3).

### 4.1 A Property of Curves in $\mathbb{R}^2$

We next describe a topological property of polygonal curves in $\mathbb{R}^2$, which will be crucial in our proof[7]. A *curve* in $\mathbb{R}^2$ is a continuous function $f : [0, \tau] \to \mathbb{R}^2$ for some $\tau \in \mathbb{R}_+$. A curve is called *polygonal* if it is piecewise linear.

For $a \in [0, \tau]$, let $f^a : [0, \tau] \to \mathbb{R}^2$ be the following curve.

$$f^a(t) = \begin{cases} f(t+a) - f(a) + f(0) & \text{if } t + a < \tau, \\ f(\tau) - f(a) + f(a+t-\tau) & \text{if } t + a \geq \tau. \end{cases}$$

---
[7]The lemma even holds for general (non-polygonal) curves. However, since we only need polygonal curves in our setting we restrict ourselves to this case since it simplifies the exposition.

Observe that $f^a(0) = f(0)$ and $f^a(\tau) = f(\tau)$ for any $a \in [0, \tau]$. The next lemma shows that any point $x$ on the segment between $f(0)$ and $f(\tau)$ is contained in some curve $f^a$.

**Lemma 7.** *Let $f : [0, \tau] \to \mathbb{R}^2$ be a polygonal curve, and let $\mu \in [0, 1]$. Then there are $a, t \in [0, \tau]$ such that $f^a(t) = \mu f(0) + (1 - \mu)f(\tau)$.*

We next give an intuitive description of the proof of the lemma: a formal proof is given in the appendix. Let $f = (f_1, f_2)$. Since the statement of the lemma is independent of changes in the coordinate system (and the claim is trivial for $f(0) = f(\tau)$), we can assume that $f(0) = (0, 0)$ and $f(\tau) = (r, 0)$ for some $r > 0$. The Gasoline Lemma [6] states that there is $a_1 \in [0, \tau]$ such that $f_2^{a_1}(t) \geq 0 \ \forall t \in [0, \tau]$. In particular, this condition is satisfied by choosing $a_1 \in \arg\min\{f_2(t) \mid t \in [0, \tau]\}$. Analogously, for $a_2 \in \arg\max\{f_2(t) \mid t \in [0, \tau]\}$, $f_2^{a_2}(t) \leq 0 \ \forall t \in [0, \tau]$. Hence, we have two curves, $f^{a_1}$ and $f^{a_2}$, one above and the other below the x-axis, both with the same endpoints $(0, 0)$ and $(r, 0)$. (See Figure 1). Furthermore, for $a$ ranging from $a_1$ to $a_2$ (in a circular sense), the curve $f^a$ continuously transforms from $f^{a_1}$ to $f^{a_2}$, always maintaining the same endpoints. Then it is intuitively clear that the union of the curves $f^a$ spans all the points on the segment from $(0, 0)$ to $(r, 0)$, hence proving the claim.

## 4.2 The Patching Procedure

In this section we describe a *patching procedure* which, given two matchings $x'$ and $x''$ and a parameter $\mu \in [0, 1]$, computes a matching $z$ satisfying $\ell^T z \leq \ell^T x_\mu$, where $x_\mu := \mu x' + (1 - \mu)x''$ is a convex combination of the first two matchings. Furthermore, the weight $w^T z$ is *close* to $w^T x_\mu$, provided that $x'$ and $x''$ have a *sufficiently* large Lagrangian weight, which is defined as follows. Let $\lambda_1^*, \lambda_2^* \in \mathbb{R}_+$ be a pair of optimal dual multipliers for the budgets in the linear program $\max\{w^T x \mid x \in P_\mathcal{M}, \ell^T x \leq L\}$. The *Langrangian weight* of $x \in [0, 1]^E$ is $\mathcal{L}(x) = w^T x - (\lambda_1^*, \lambda_2^*)(\ell^T x - L)$. Notice, that by the theory of Lagrangian duality we have $w^* = \max\{\mathcal{L}(x) \mid x \in P_\mathcal{M}\}$, where $w^*$ is the weight of an optimal LP solution, i.e., $w^* = \max\{w^T x \mid x \in P_\mathcal{M}, \ell^T x \leq L\}$ (see [19] for more information on Lagrangian duality).

We need the following notion of almost matching.

**Definition 8.** *For $r \in \mathbb{N}$, an $r$-almost matching in $G$ is a (possibly fractional) vector $y \in [0, 1]^E$ such that it is possible to set at most $r$ components of $y$ to zero to obtain a matching.*

We denote by $\mathcal{M}_r$ the set of all $r$-almost matchings in $G$. Given an $r$-almost matching $y$, we let a *corresponding matching* $z \in \mathcal{M}$ be a matching obtained by setting to zero the fractional components of $y$, and then computing a maximal matching in the resulting set of edges (in particular, we might need to set to 0 some 1 entries of $y$ to obtain $z$). Notice that $w^T z \geq w^T y - r w_{\max}$, where $w_{\max}$ is the largest weight.

Our patching procedure first constructs a 2-almost matching $y$, and then returns a corresponding matching $z$. We next show how to compute $y$. Let us restrict our attention to the following set of candidate 2-almost matchings. Recall that $s = x' \Delta x''$ is a set of paths and cycles. We construct an auxiliary graph $C$, consisting of one cycle $(e_0, e_1, \ldots, e_{\tau-1})$, with the following property: there is a bijective mapping between the edges of $C$ and the edges of $s$ such that two consecutive edges of $C$ are either consecutive in some path/cycle or belong to different paths/cycles. This can be easily achieved by cutting each cycle, appending the resulting set of paths one to the other, and gluing together the endpoints of the obtained path. For $t \in [0, \tau]$, we define $s(t) \in [0, 1]^E$ as

$$(s(t))(e) = \begin{cases} 1 & \text{if } e = e_i, \ i < \lfloor t \rfloor; \\ t - \lfloor t \rfloor & \text{if } e = e_i, \ i = \lfloor t \rfloor; \\ 0 & \text{otherwise.} \end{cases}$$

Moreover, for $a, t \in [0, \tau]$, we define

$$[0,1]^E \ni s^a(t) = \begin{cases} s(a+t) - s(a) & \text{if } a+t \leq \tau; \\ s(a+t-\tau) + s(\tau) - s(a) & \text{if } a+t > \tau. \end{cases}$$

Intuitively, $a$ and $(a+t) \pmod{\tau}$ define a (fractional) subpath of $C$, and $s^a(t)$ is the (fractional) incidence vector corresponding to that subpath. Eventually we define

$$y^a(t) := x' \triangle s^a(t).$$

Note that $y^a(t)$ is equal to $x'$ and $x''$ for $t = 0$ and $t = \tau$, respectively.

**Lemma 9.** *For any $a, t \in [0, \tau]$, $y^a(t)$ is a 2-almost matching.*

*Proof.* One can easily observe that a matching can be obtained by setting the two components of $y^a(t)$ to zero that correspond to the edges $e_{\lfloor a \rfloor}$ and $e_{\lfloor (a+t) \pmod{\tau} \rfloor}$. □

The following lemma shows that, in polynomial time, one can find a 2-almost matching $y$ with lengths $\ell^T y$ equal to the lengths of any convex combination of the two matchings $x'$ and $x''$.

**Lemma 10.** *Let $\mu \in [0, 1]$ and $x_\mu = \mu x' + (1-\mu)x''$. In polynomial time, $a, t \in [0, \tau]$ can be determined such that $\ell^T y^a(t) = \ell^T x_\mu$.*

*Proof.* Let $f : [0, \tau] \to \mathbb{R}^2$ be the polygonal curve defined by $f(t) = \ell^T y^0(t)$. Since $f(0) = \ell^T x'$ and $f(\tau) = \ell^T x''$, we have by Lemma 7 that there exists $a, t \in [0, \tau]$ such that $f^a(t) = \ell^T x_\mu$. Since $f^a(t) = \ell^T y^a(t)$, $y := y^a(t)$ satisfies the claim.

The values of $\lfloor a \rfloor$ and $\lfloor a+t \rfloor$ can be guessed in polynomial time by considering $O(n^2)$ possibilities. Given those two rounded values, the actual values of $a$ and $t$ can be obtained by solving a linear program with a constant number of variables and constraints. □

Our patching procedure simply computes a 2-almost matching $y = y^a(t)$ with $\ell^T y = \ell^T x_\mu$, exploiting the lemma above, and then returns a corresponding matching $z$, by applying the procedure explained in the proof of Lemma 9. Trivially, $\ell^T z \leq \ell^T y = \ell^T x_\mu$. We next show that, if $x'$ and $x''$ have sufficiently large Lagrangian weight, then the weight of $z$ is *close* to the weight of $x_\mu$.

**Lemma 11.** *Assume $\mathcal{L}(x') \geq w^* - \Gamma$ and $\mathcal{L}(x'') \geq w^* - \Gamma$ for some $\Gamma \in \mathbb{R}_+$. Then the matching $z$ returned by the patching procedure satisfies $w^T z \geq w^T x_\mu - 2w_{\max} - \Gamma$ and $\ell^T z \leq \ell^T x_\mu$.*

*Proof.* By Lemma 10 we have $\ell^T y = \ell^T x_\mu$, and since $z \leq y$, we get $\ell^T z \leq \ell^T x_\mu$. Let $\overline{x}_\mu = x' + x'' - x_\mu = (1-\mu)x' + \mu x''$. Since $\mathcal{L}(x') \geq w^* - \Gamma$, $\mathcal{L}(x'') \geq w^* - \Gamma$ and $\mathcal{L}$ is linear, we have $\mathcal{L}(x_\mu) \geq w^* - \Gamma$ and $\mathcal{L}(\overline{x}_\mu) \geq w^* - \Gamma$. Recall that $y = y^a(t)$ for a proper choice of $a, t \in [0, \tau]$. Let $\overline{y} := x' + x'' - y$. Notice that $\overline{y} = y^{a'}(\tau - t)$ where $a' = (a+t) \pmod{\tau}$, and hence, $\overline{y}$ is also a 2-almost matching by Lemma 9. Let $\overline{z}$ be the matching corresponding to $\overline{y}$ obtained by applying the procedure explained in the proof of Lemma 9 to $\overline{y}$. Notice that the pairs $(z, y)$ and $(\overline{z}, \overline{y})$ differ on the same two (or less) components. Hence

$$w^T z + w^T \overline{z} + 2w_{\max} \geq w^T y + w^T \overline{y} = w^T x_\mu + w^T \overline{x}_\mu. \qquad (4)$$

Since $y + \overline{y} = x_\mu + \overline{x}_\mu$ and $\ell^T y = \ell^T x_\mu$, we get $\ell^T \overline{y} = \ell^T \overline{x}_\mu$. Thus, $\ell^T \overline{z} \leq \ell^T \overline{x}_\mu$ since $\overline{z} \leq \overline{y}$. This can be rewritten as $\mathcal{L}(\overline{z}) - w^T \overline{z} \geq \mathcal{L}(\overline{x}_\mu) - w^T \overline{x}_\mu$. Since $\mathcal{L}(\overline{x}_\mu) \geq w^* - \Gamma$ and $\mathcal{L}(\overline{z}) \leq w^*$, we obtain $w^T \overline{z} \leq w^T \overline{x}_\mu + \Gamma$. Combining this result with (4) implies $w^T z \geq w^T x_\mu - 2w_{\max} - \Gamma$. □

### 4.3 The Algorithm

Our PTAS works as follows. Initially it guesses the $6/\varepsilon$ heaviest edges $E_H$ in the optimum solution, and reduces the problem consequently. Then it computes a vertex $x^* \in P_\mathcal{M}$ of the polytope $\{x \in P_\mathcal{M} \mid \ell^T x \le L\}$ of maximum weight $w^* := w^T x^*$. As $x^*$ is a vertex solution of the polytope $P_\mathcal{M}$ with two additional constraints, it lies on a face of $P_\mathcal{M}$ of dimension at most two. Hence, by Carathéodory's Theorem, $x^*$ can be expressed as a convex combination $x^* = \alpha_1 x_1 + \alpha_2 x_2 + \alpha_3 x_3$ of three matchings $x_1, x_2, x_3 \in P_\mathcal{M}$. Let $\mu' = \alpha_1/(\alpha_1 + \alpha_2)$ and $\mu'' = (\alpha_1 + \alpha_2)/(\alpha_1 + \alpha_2 + \alpha_3)$. Applying Lemma 11 to $x_1$ and $x_2$ with $\mu = \mu'$, a matching $z'$ is obtained. Applying Lemma 11 to $z'$ and $x_3$ with $\mu = \mu''$, we obtain a matching $z''$. The algorithm returns $z''$ plus $E_H$.

*Proof of Theorem 6.* Consider the algorithm above. The initial guessing can be performed in $O(|E|^{6/\varepsilon})$ time. Since it is possible to efficiently separate over $P_\mathcal{M}$, $x^*$ can be computed in polynomial time [31]. The same holds for the decomposition of $x^*$ into three matchings by standard techniques (see for example [30]). Lemma 10 implies that the patching can be done in polynomial time. Hence the proposed algorithm runs in polynomial time as claimed.

Since $\mathcal{L}(x^*) = w^*$ and $\mathcal{L}(x) \le w^*$ for $x \in P_\mathcal{M}$, we get $\mathcal{L}(x_1) = \mathcal{L}(x_2) = \mathcal{L}(x_3) = w^*$. Let $u := \mu' x_1 + (1 - \mu') x_2$ and $v := \mu'' z' + (1 - \mu'') x_3$. By Lemma 11, matching $z'$ satisfies $\ell^T z' \le \ell^T u$ and $w^T z' \ge w^T u - 2 w_{\max}$. Since $u$ is a convex combination of $x_1$ and $x_2$, we have $\mathcal{L}(u) = w^*$. Furthermore, by the relations between the lengths and weight of $z'$ and $u$, we get $\mathcal{L}(z') \ge \mathcal{L}(u) - 2w_{\max} = w^* - 2w_{\max}$.

By Lemma 11, matching $z''$ satisfies $\ell^T z'' \le \ell^T v$ and $w^T z'' \ge w^T v - 4 w_{\max}$. We observe that $z''$ satisfies the budget constraints since

$$\ell^T z'' \le \ell^T v = \ell^T((\alpha_1 + \alpha_2) z' + \alpha_3 x_3) \le \ell^T((\alpha_1 + \alpha_2) u + \alpha_3 x_3) = \ell^T x^* \le L.$$

Furthermore,

$$w^T z'' \ge w^T v - 4 w_{\max} = w^T((\alpha_1 + \alpha_2) z' + \alpha_3 x_3) - 4 w_{\max}$$
$$\ge w^T((\alpha_1 + \alpha_2) u + \alpha_3 x_3) - 6 w_{\max} = w^* - 6 w_{\max}.$$

Let $OPT'$ be an optimum solution to the reduced problem. Of course, $w^* \ge w(OPT')$. Furthermore, the weight of the guessed edges $E_H$ is at least $6/\varepsilon \, w_{\max}$. Since $w(OPT) = w(E_H) + w(OPT')$, we can conclude that the solution returned by the algorithm has weight at least $w(E_H)(1 - \varepsilon) + w(OPT') \ge (1 - \varepsilon) w(OPT)$. $\square$

## 5 Conclusions

A first obvious open problem is finding a PTAS for the $k$-BUDGETED MATCHING problem for any $k = O(1)$. It is interesting to notice that most parts of the approach presented in Section 4 can easily be generalized to an arbitrary constant number of budget constraints. More precisely, the only part that is tailored to two budgets is Lemma 7, which is only valid for curves in two dimensions. We believe that for every constant $k \in \mathbb{N}$ there is $r(k) \in \mathbb{N}$ such that the following generalized version of Lemma 7 holds.

**Conjecture 12.** *Let $f : [0, \tau] \to \mathbb{R}^k$ be a curve, and let $\mu \in [0, \tau]$. Then there are $r(k)$ disjoint intervals $[a_1, b_1], \ldots, [a_{r(k)}, b_{r(k)}] \subseteq [0, \tau]$ such that $f(0) + \sum_{i=1}^{r(k)} (f(b_i) - f(a_i)) = \mu f(0) + (1 - \mu) f(\tau)$.*

In particular, for $k = 2$, the pair $(a, t)$ given by Lemma 7 defines either one interval $[a, a+t]$ or two intervals $[a, \tau]$ and $[0, a + t - \tau]$ satisfying the claim above. If Conjecture 12 holds, then a PTAS for the $k$-budgeted case can be obtained by following a procedure analogous to the one that we presented here.

Another interesting direction for further research, is to check whether a similar technique can be applied to $k$-BUDGETED MATROID INTERSECTION INDEPENDENT SET. A crucial property that we exploited in this paper is the well-known fact that a matching can be transformed to another matching in the same graph by exchanging edges on alternating paths and cycles. Similar exchange properties are known for the intersection of matroids. However, they do not seem to allow for an easy adaption of the presented algorithm.

Another set of questions involves the problems considered here but with one budget. Is there a fully-polynomial PTAS (FPTAS) for 1-BUDGETED SPANNING TREE and 1-BUDGETED MATCHING? We remark that, as noted in [6], an FPTAS for the second problem would imply a deterministic algorithm for EXACT MATCHING with polynomial weights, which is a long-standing open problem. In the case of 1-BUDGETED PERFECT MATCHING, a PTAS is not known.

**Acknowledgements.** The authors wish to thank Friedrich Eisenbrand for supporting their visit to EPFL and for helpful discussions.

**Figure 1** The full line indicates $f$, and the dashed lines $f^{a_1}$ and $f^{a_2}$, respectively.

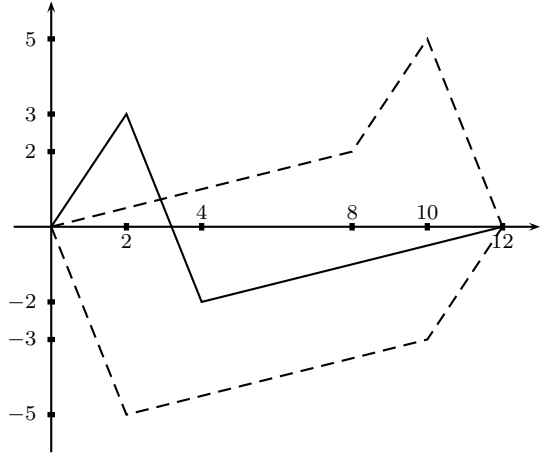

# Appendix

*Proof of Theorem 1.* We show that deciding feasibility of the considered problems is $NP$-complete. It is sufficient to prove the claim for $k = 2$. Consider first 2-BUDGETED SPANNING TREE: the claim for $k$-BUDGETED MATROID BASIS and, consequently, for $k$-BUDGETED MATROID INTERSECTION BASIS trivially follows. Let $\mathcal{P}^+$ denote the problem, and $\mathcal{P}^\pm$ its variant with arbitrary (i.e., positive and/or negative) lengths. Of course, $\mathcal{P}^\pm$ includes $\mathcal{P}^+$ as a special case. To see the opposite reduction, observe that a spanning tree contains exactly $n-1$ edges. Hence, by adding a sufficiently large value $M$ to all the lengths, and adding $(n-1)M$ to the budgets, one obtains an equivalent problem with non-negative lengths. It is easy to see that $\mathcal{P}^\pm$ includes as a special case the problem $\mathcal{P}^=$ of determining, for a given length function $\ell'(\cdot)$ and target $L'$, whether there exists a spanning tree $T$ of length $\ell'(S) = L'$: a reduction is obtained by setting $\ell_1(\cdot) = -\ell_2(\cdot) = \ell'(\cdot)$ and $L_1 = -L_2 = L'$. Hence it is sufficient to show that $\mathcal{P}^=$ is NP-complete. We do that via the following reduction from PARTITION: given $\alpha_1, \alpha_2, \ldots, \alpha_q \in \mathbb{Q}$ and a target $A \in \mathbb{Q}$, determine whether there exists a subset of $\alpha_i$'s of total value $A$. Consider graph $G_q$, consisting of $q$ cycles $C_1, C_2, \ldots, C_q$, with $C_i = (a_i, b_i, c_i, d_i)$ and $c_i = a_{i+1}$ for $i = 1, 2, \ldots, q-1$. Let $\ell'(a_i b_i) = \alpha_i$, $i = 1, 2, \ldots, k$, and set to zero all the other lengths. The target is $L' = A$. Trivially, for each spanning tree $T$ and each cycle $C_i$, the length of $T \cap C_i$ is either $0$ or $\alpha_i$. Hence, the answer to the input partition problem is yes if and only if the same holds for the associated instance of $\mathcal{P}^=$.

Consider now 2-BUDGETED PERFECT MATCHING. Since each perfect matching contains exactly $n/2$ edges, with the same argument and notation as above it is sufficient to prove the $NP$-completeness of the problem $\mathcal{P}^=$ of determining, for a given length function $\ell'(\cdot)$ and target $L'$, whether there exists a perfect matching $M$ of length $\ell'(M) = L'$. We use a similar reduction from PARTITION as above. The graph is again given by the cycles $C_1, \ldots, C_q$. However, this time each cycles forms its own connected component. Furthermore, the lengths are given by $\ell'(a_i b_i) = \ell'(c_i d_i) = \alpha_i$, $i = 1, 2, \ldots, k$, all the other lengths are zero, and $L' = 2A$. It is easy to see that, for each perfect matching $M$ and each cycle $C_i$, the length of $T \cap C_i$ is either $0$ or $2\alpha_i$. The claim follows.

Eventually consider 2-BUDGETED SHORTEST PATH. We restrict our attention to the graph $G_q$ as used for the spanning tree reduction, and let $(s, t) = (a_1, c_q)$. Since any $s$-$t$ path in this graph uses exactly $2q$ edges, we have by the usual argument that it is sufficient to show the $NP$-completeness of the problem $\mathcal{P}^=$ of determining, for a given length function $\ell'(\cdot)$ and target L', whether there exists an *s-t* path $P$ of length $\ell'(P) = L'$. The claim follows by essentially the same reduction as in the spanning tree case. $\square$

*Proof of Lemma 7.* Without loss of generality, we can assume that $f = (f_1, f_2)$ satisfies.

(i) $f(0) = (0,0)$ and $f(\tau) = (r, 0)$ for $r > 0$.

(ii) $f_2(t) \geq 0 \ \forall t \in [0, \tau]$,

(iii) $f$ is not self-intersecting (i.e., $f$ is an injection).

Let $f = (f_1, f_2)$. Since the statement of the lemma is independent of changes in the coordinate system (and the claim is trivial for $f(0) = f(\tau)$), we can assume that $f(0) = (0, 0)$ and $f(\tau) = (r, 0)$ for some $r > 0$. Hence, (i) holds. The Gasoline Lemma [6] states that there is $\bar{a} \in [0, \tau]$ such that $f_2^{\bar{a}}(t) \geq 0 \ \forall t \in [0, \tau]$. More precisely, this condition is satisfied by choosing $\bar{a} \in \arg\min\{f_2(t) \mid t \in [0, \tau]\}$. One can easily observe that if the lemma is true for $f^{\bar{a}}$ then it also holds for $f$. Hence, we can assume, by replacing $f$ by $f^{\bar{a}}$, that Property (ii) holds. Property (iii) can be enforced by removing loops.

Furthermore, we assume $\mu \in (0, 1)$, otherwise the claim is trivially true. Let $v = (v_1, 0) = \mu f(0) + (1 - \mu) f(\tau) = (1 - \mu) f(\tau)$ and $g : [0, \tau] \to \mathbb{R}^2$ be the translation of $f$ by $v$:

$$g(t) = f(t) + v \quad \forall t \in [0, \tau].$$

Let $u = \max\{f_2(t) \mid t \in [0, \tau]\}$ and $p = \min\{t \in [0, \tau] \mid f_2(t) = u\}$, i.e., $f_2$ attains its maximum the first time at $p$. Let $f' : [p, \tau] \to \mathbb{R}^2$ with $f'(t) = f(t)$ be the subcurve of $f$ over the interval $[p, \tau]$, and let $g' : [0, p] \to \mathbb{R}^2$ with $g'(t) = g(t)$ be the subcurve of $g$ over the interval $[0, p]$. In the following we show that $f'$ and $g'$ intersect. Consider the endpoints of $f'$ and $g'$. The endpoints $f'(\tau) = (r, 0)$ and $g'(0) = (v_1, 0)$ both lie on the x-axis, and since $v_1 = (1 - \mu) r < r$, $f'(\tau)$ lies to the right of $g'(0)$. Similarly, the other two endpoints $f'(p) = (f_1(p), u)$ and $g'(p) = (f_1(p) + v_1, u)$ have the same y-components, where this time $f'(p)$ is to the left of $g'(p)$. Since the second component of both curves lies between 0 and $u$, one can easily deduce that they have to cross. In more detail, one way to show this is to consider the polygonal curve $h$ joining, in the given order, $f'(p) = (f_1(p), u)$, $(-M, u)$, $(-M, -1)$, $(r, -1)$ and $(r, 0) = f'(\tau)$, where $M > 0$ is a large value such that $h$ does not intersect $g'$. The concatenation of $h$ with $f'$ forms a closed, not self-intersecting curve, which - by Jordan's Curve Theorem [22] - divides $\mathbb{R}^2$ into two regions, a bounded one and an unbounded one. Furthermore, $g'$ has one endpoint in one region and the other endpoint in the other region. Again by Jordan's curve theorem, we have that $g'$ and $f'$ intersect (since we have by construction that $g'$ does not intersect $h$). Hence, there exists $t_1 \in [p, \tau]$ and $t_2 \in [0, p]$ such that $f(t_1) = g(t_2)$. Since $f(p) \neq g(p)$, we have $t_2 < t_1$. The claim is satisfied by

$$f^{t_2}(t_1 - t_2) = f(t_1 - t_2 + t_2) - f(t_2) = g(t_2) - f(t_2) = v.$$

$\square$